\documentclass[%
 aip,
 amsmath,amssymb,
preprint,
]{revtex4-1}

\usepackage{graphicx}
\usepackage{dcolumn}
\usepackage{bm}

\usepackage[utf8]{inputenc}
\usepackage[T1]{fontenc}
\usepackage{mathptmx}

\usepackage{color}

\begin{document}

\preprint{}

\title[]{Stress-controlled zero-field spin splitting in silicon carbide}

\author{I.~D.~Breev}
  \email{breev.ilia.d@mail.ioffe.ru}

\author{A.~V.~Poshakinskiy}
\author{V.~V.~Yakovleva}
\author{S.~S.~Nagalyuk}
\author{E.~N.~Mokhov}

\affiliation{Ioffe Institute, Russia, Saint-Petersburg, Politechnicheskaya st., 26}

\author{R.~H\"ubner}
\affiliation{Institute of Ion Beam Physics and Materials Research,
Helmholtz-Zentrum Dresden-Rossendorf (HZDR), Germany, Dresden, Bautzner Landstr. 400}
\author{G.~V.~Astakhov}
\affiliation{Institute of Ion Beam Physics and Materials Research,
Helmholtz-Zentrum Dresden-Rossendorf (HZDR), Germany, Dresden, Bautzner Landstr. 400}
\affiliation{Ioffe Institute, Russia, Saint-Petersburg, Politechnicheskaya st., 26}

\author{P.~G.~Baranov}
\homepage{http://www.ioffe.ru/labmsc/ru/main.html}

\author{A.~N.~Anisimov}
\email{aan0100@gmail.com}
\affiliation{Ioffe Institute, Russia, Saint-Petersburg, Politechnicheskaya st., 26}

\date{\today}

\begin{abstract}
We report the influence of static mechanical deformation on the zero-field splitting of silicon vacancies in silicon carbide at room temperature.  We use AlN/6H-SiC heterostructures deformed by growth conditions and monitor the stress distribution as a function of distance from the heterointerface with spatially-resolved confocal Raman spectroscopy. The zero-field splitting of the V1/V3 and V2 centers in 6H-SiC, measured by optically-detected magnetic resonance, reveal significant changes at the heterointerface  compared to the bulk value. This approach allows unambiguous determination of the spin-deformation interaction constant, which turns out to be $0.75 \, \mathrm{GHz}$ for the V1/V3 centers and $0.5 \, \mathrm{GHz}$ for the V2 centers. Provided piezoelectricity of AlN, our results offer a strategy to realize the on-demand fine tuning  of spin transition energies in SiC by deformation.
\end{abstract}

\maketitle

Atomic-scale defects in bulk and nanocrystalline silicon carbide (SiC) are promising for quantum information processing \cite{quantumproc,Castelletto:2020kr,Son:2020kh}, nanophotonics \cite{lukin2020integrated},   and sensing\cite{Tarasenko:2017ky}. The breakthrough quantum properties have recently been discovered for a family of uniaxially oriented silicon-vacancy-related color centres in the ground and excited states of the hexagonal (4H-SiC, 6H-SiC) and rhombic (15R-SiC) polytypes \cite{Baranov:2011ib, PhysRevLett, Riedel:2012jq, Falk:2013jq, PhysRevX}. Silicon vacancies ($\mathrm{V_{Si}}$) in SiC are of particular interest, because they possess half-integer spin ($S=3/2$) \cite{Kraus:2013di}, which can be optically polarized and read out even at elevated temperatures up to  250$^{\circ}$C by means of a standard optically-detected magnetic resonance (ODMR) technique \cite{Kraus:2013vf}. They show long coherence time \cite{Simin:2017iw} allowing highly-sensitive quantum magnetometry \cite{PhysRevX, JJETPLett, Soltamov:2019hr} and thermometry \cite{Anisimov:2016er, JJETPLett}. Furthermore, isolated single $\mathrm{V_{Si}}$ centers \cite{Widmann:2014ve, Fuchs:2015ii} reveal high readout contrast \cite{Baranov:2011ib, Nagy:2018ey} and high spectral stability \cite{Nagy:2019fw}  providing a basis for the implementation of quantum repeaters \cite{Morioka:2020iv, Lukin:2019fh}. 

The fine tuning of the spin and optical properties of the color centers in SiC is crucial for quantum sensing and communication. The most frequent approach is based on the Stark shift of the zero-phonon line and zero-field spin splitting \cite{Falk:2014fh, Ruhl:2019hs, Lukin:2020jb}. Alternatively, resonant interaction with acoustic waves can be used \cite{Whiteley:2019eu, HernandezMinguez:2020kv}. Here, we demonstrate that mechanical stress modifies the zero-field splitting of the $\mathrm{V_{Si}}$ centers in SiC. To this end, we investigate AlN/6H-SiC heterostructures, where the mechanical stress is induced at the heterointerface due to different lattice parameters of AlN and 6H-SiC and relaxes away from the heterointerface. Using confocal Raman spectroscopy and the ODMR technique, we simultaneously monitor the  lattice stress through  the shift of the 6H-SiC Raman modes and the zero-field spin splitting $2D$ through the measurement  of the spin resonance frequency. We concentrate on the V1/V3 and V2 $\mathrm{V_{Si}}$ centers in 6H-SiC, associated with different crystallographic sites, for which the zero-field splitting constant $D$ is expected to have opposite signs\cite{PhysRevB.98.195204}. Using this approach, we determine the spin-deformation interaction constant and confirm the opposite sign of $D$ in these centers. Given that AlN is a pizoelectric material, which is also used for the fabrication of electro-acoustical transducers on SiC \cite{Whiteley:2019eu}, our results provide a basis for the controllable and local tuning of the zero-field spin splitting in $\mathrm{V_{Si}}$ and other spin centers.  



\begin{figure}[b]
\includegraphics[width=.95\linewidth]{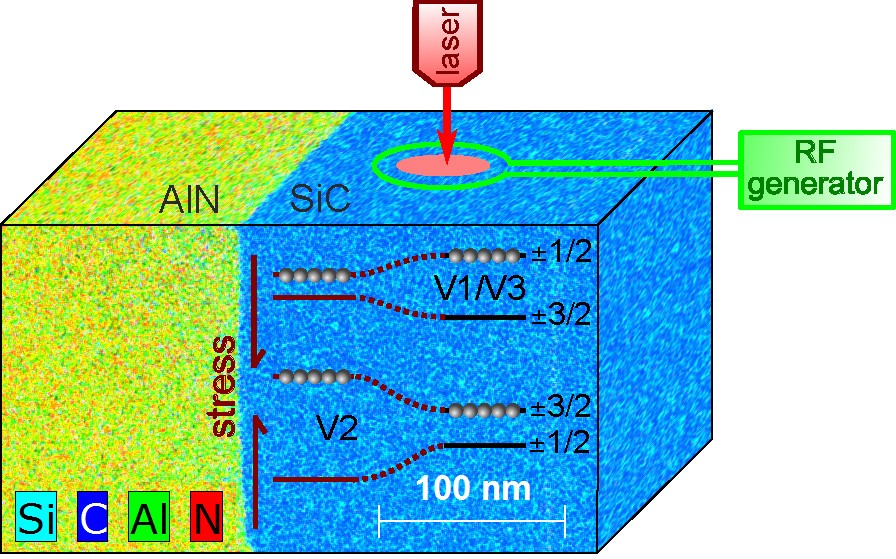}
\caption{\label{Figure 1} Color-coded chemical element distributions obtained by EDXS across the AlN/6H-SiC heterointerface. The lines schematically represent the spatial distribution (not to scale) of the zero-field splitting in the V1/V3 and V2 $\mathrm{V_{Si}}$ centers in 6H-SiC near the heterointerface and in the bulk. The solid circles show the preferentially populated spin states under optical pumping. A simplified scheme of the ODMR registration shows an excitation laser (523~nm or 785~nm) and a tunable RF generator. The laser spot is not to scale.   }
\end{figure} 
\begin{figure*}
\includegraphics{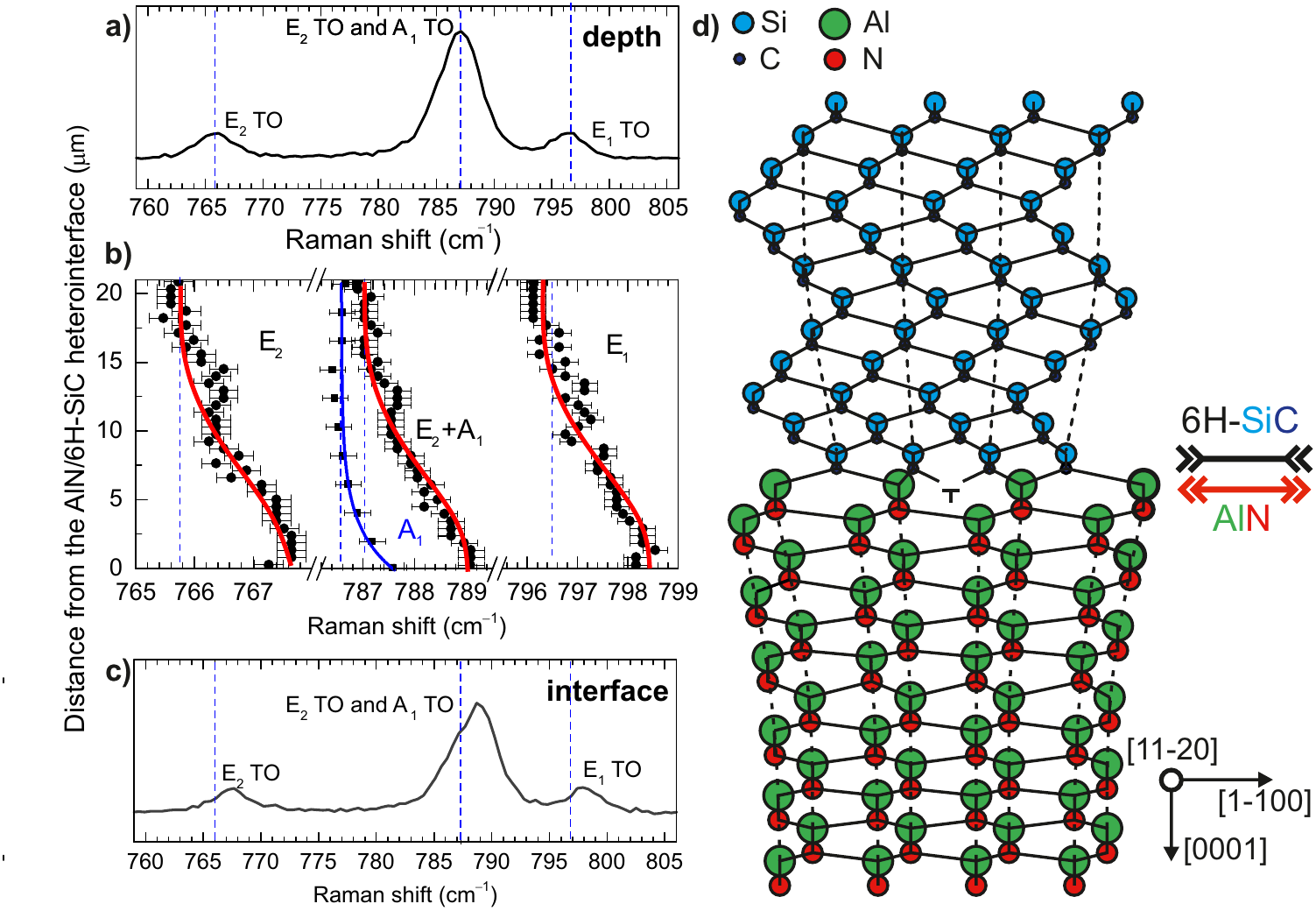}
\caption{\label{Figure 2}  (a) 6H-SiC Raman spectrum at a distance of $20 \, \mathrm{\mu m}$ from the AlN/6H-SiC heterointerface. (b) Symbols are the position of the Raman peaks as a function of the distance from the heterointerface.
The solid red and blue lines are fits. 
The blue line corresponds to the Raman peak shift $A_1$ registered in another optical polarisation. (c) 6H-SiC Raman spectrum at the AlN/6H-SiC heterointerface. The vertical dashed lines show the position of the Raman modes in  stress-free 6H-SiC. (d) Schematic presentation of the lattice deformation in AlN and 6H-SiC due to the stress variation across the heterointerface.}
\end{figure*}

The AlN/6H-SiC heterostructures are grown by the sublimation sandwich method \cite{CrystalGrowth}. First, a 6H-SiC substrate is grown at a temperature of 2200$^{\circ}$C at an argon pressure in the range of 5 to 50 torr. The crystal growth rate under these conditions is about 100~$\mathrm{\mu m/h}$. The intrinsic impurity concentrations obtained by secondary ion mass spectroscopy (SIMS) are  N = $7.3 \times10^{16} \, \mathrm{cm^{-2}}$,  B = $4.4 \times 10^{16} \, \mathrm{cm^{-2}}$, and  Al = $4.2 \times 10^{16} \, \mathrm{cm^{-2}}$. To prepare the 6H-SiC substrate for the AlN growth, it is polished and diced in chips.  AlN crystals are grown under the following basic parameters: the growth cell temperature is about 2000$^{\circ}$C, the gap between the source and the seed is 3-10~mm, the temperature gradient in the growth zone is about 5K/mm, the Nitrogen pressure in the growth chamber is 0.3-1~atm. The growth chamber is preliminarily heated in vacuum to 1600$^{\circ}$C and then in nitrogen atmosphere to 1800$^{\circ}$C. The maintaining of these parameters ensures the removal of oxygen and the formation of a thin AlN buffer layer. Main part of the growth process takes place at a temperature of about 2000$^{\circ}$C. In all cases, we use a resistive graphite heater.

To create the $\mathrm{V_{Si}}$ centers, we perform electron irradiation with an energy of $2 \, \mathrm{MeV}$ to a fluence of $10^{18} \, \mathrm{cm^{-2}}$, providing a homogeneous $\mathrm{V_{Si}}$ distribution and the optimal ODMR signal-to-noise ratio  \cite{PhysRevApp}. After irradiation, the crystals are cleaved perpendicular to the growth surface and along the dislocation zone in AlN. The cleavage surfaces are optically smooth, allowing confocal optical scans across the AlN/6H-SiC heterointerface, as schematically shown in Fig.~\ref{Figure 1}.  Chemical element mapping based on energy-dispersive X-ray spectroscopy (EDXS) in scanning transmission electron microscopy mode shows a sharp heterointerface and the absence of any kind of solid solution.

To measure Raman, photolumenescence (PL), and ODMR spectra, we use a home-made spectrometer based on the confocal microscope ("Spectra", NT-MDT SI), as schematically shown in Fig.~\ref{Figure 1}. It is used in two modes ensuring signal collection from the same volume. A 100X objective with NA = 0.9 and a 100-$\mathrm{\mu m}$ pinhole provide a collection volume of about $ 5 \, \mathrm{\mu m^3}$. In one mode, Raman and PL spectra are excited with a semiconductor laser ($\lambda$ = 532~nm, excitation power ca. 10~mW) and recorded with a SolarTII monochromator and an Andor CCD camera. In the second mode, ODMR spectra are recorded with a 785-nm laser and a NIR detector (Hamamtsu, C12483-250). The radiofrequency (RF) from a signal generator is modulated at 420~Hz and sent to an antenna placed in the vicinity of the sample. The RF-induced changes in the PL signal detected by the photodiode are locked-in at the modulation frequency of 420~Hz. The ODMR spectra are recorded by varying RF in the range from 1 to 160~MHz. To reduce magnetic noise, we compensate the external magnetic field along the $\mathrm{V_{Si}}$ symmetry axis to a level better than $100 \, \mathrm{\mu T}$. 


First, we analyze the evolution of the 6H-SiC Raman spectra depending on the distance from the AlN/6H-SiC heterointerface (Fig.~\ref{Figure 2}). Figures~\ref{Figure 2}(a) and (c) show three Raman peaks associated with the $E_2$ TO, $E_2$ TO and $A_1$ TO, and $E_1$ TO phonon modes in 6H-SiC \cite{SiCRaman,oldRaman} at a distance of $20 \, \mathrm{\mu m}$ and exactly at the heterointerface, respectively. We extract the spectral positions of these peaks from their maxima averaged by 4 points within 1 $\mu$m along the perpendicular direction to the scanning  axis with an accuracy of $0.2  \, \mathrm{cm^{-1}}$. The two peaks corresponding to the $E_2$ TO and $A_1$ TO modes at around 787\,cm$^{-1}$ are hardly resolved, so we had to use different Raman geometries to separate them, as described in Ref.~\onlinecite{breev2020stress}.  The spectral shifts of the Raman peaks are summarized in Fig.~\ref{Figure 2}(b). 
At a distance of about $15\,\mathrm{\mu m}$ from the interface, the Raman lines coincide with the values of an undeformed crystal, which is visualized by the red and blue fitting curves. The shift of the Raman frequencies at the heterointerface deviates from the bulk values. In particular, $\Delta\textit{w}_{E_2}=2.1$\,cm$^{-1}$ for the ${E_2}$ modes at around 767\,cm$^{-1}$ and at 787\,cm$^{-1}$, $\Delta \textit{w}_{A_1}=1.0$\,cm$^{-1}$ for the ${A_1}$ mode at around 787\,cm$^{-1}$, and $\Delta \textit{w}_{E_1}=2.4$\,cm$^{-1}$ for the ${E_1}$ mode at around 797\,cm$^{-1}$. 

The positions of the Raman peaks at the heterointerface are different from those in the bulk due the deformation of the 6H-SiC crystal in the vicinity of the heterointerface\cite{breev2020stress}. This deformation is characterized by the stress components along and perpendicular to the c-axis, $\sigma_{\|}$ and $\sigma_{\perp}$, respectively. The spectral shift of the Raman modes is given by $\Delta \textit{w}_i = 2\textit{a}'_i \sigma_{\perp} + \textit{b}'_i \sigma_{\|} $. Using the values $\textit{a}'_{A_1}=-0.46$\,cm$^{-1}$/GPa, $\textit{b}'_{A_1} = -2.67$\,cm$^{-1}$/GPa and $\textit{a}'_{E_2}=-1.55$\,cm$^{-1}$/GPa, $\textit{b}'_{E_2} = -0.74$\,cm$^{-1}$/GPa  for the $A_1$ and $E_2$ modes\cite{Sugie}, we calculate that the stress components are $\sigma_{\perp} = -0.64 \, \mathrm{GPa}$ and $\sigma_{\|} = -0.15 \, \mathrm{GPa}$. The corresponding deformations are calculated using the elastic constants \cite{Ctensor} and read $\textit{u}_{\perp} = -1.0 \times 10^{-3} $ and $\textit{u}_{\|} = -0.08 \times10^{-4} $. This result indicates that the heterointerface induces mostly the stress in its plane, i.e., $\textit{u}_{\perp} \gg \textit{u}_{\|}$, which agrees well with previous studies\cite{Liu}. The 6H-SiC crystal is compressively stressed  at the heterointerface, as schematically depicted in Fig.~\ref{Figure 2}(d). The detailed analysis shows that the stress has arised when the structure was cooled from growth to room temperature due to the different coefficients of thermal expansion of SiC and AlN\cite{breev2020stress}.

Based on the obtained deformation at the heterointerface, we now analyze its effect on the spin centers. An effective spin Hamiltonian in the presence of deformation and in the absence of external magnetic field reads\cite{Poshakinskiy:2019bi,Udvarhelyi:2018tg} 
\begin{equation}
\label{Hamiltonian}
    H = D\left(S_z^2-\frac{3}{4}\right) + \Xi\sum_{\alpha\beta}u_{\alpha\beta}S_\alpha S_\beta .
\end{equation}
Here,  \textit{D} is the zero-field splitting constant and $u_{\alpha\beta}$ is the deformation tensor. Hamiltonian (\ref{Hamiltonian}) accounts for the deformation in the spherical approximation, when its effect on the spin center is described by a single constant $\Xi$. In the case of uniaxial deformation along the z-axis ($u_{zz} = u_\|, u_{xx} = u_{yy} = u_\perp$), Hamiltonian (\ref{Hamiltonian}) describes the strain-induced modification of the zero-field splitting constant,
\begin{equation}
\label{Variation}
    \delta D = \Xi (u_\| - u_\perp),
\end{equation}
Such a variation is directly accessible with the ODMR technique. 

\begin{figure}
\includegraphics{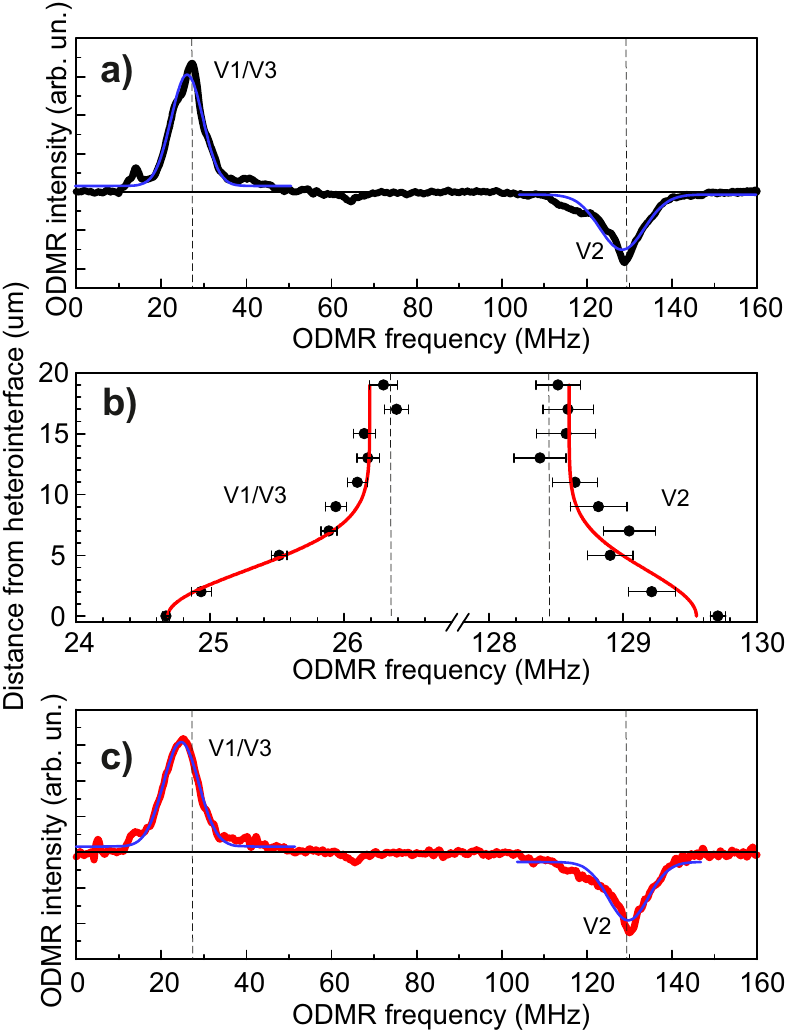}
\caption{\label{Figure 3} (a) 6H-SiC ODMR spectrum at a distance of $20 \, \mathrm{\mu m}$ from the AlN/6H-SiC heterointerface. (b) Symbols are the position of the ODMR peaks as a function of the distance from the heterointerface. The V1/V3 and V2 ODMR peaks are shifted by $-1.5$ and $1.0 \, \mathrm{MHz}$,  respectively.  The red lines are fits to a Gaussian with HWHM of $4.5 \, \mathrm{\mu m}$.(c) 6H-SiC ODMR spectrum on AlN/6H-SiC heterointerface.  The vertical dashed lines show the position of the ODMR peaks in stress-free 6H-SiC. The blue line in (on (a) and (c)) the graph shows the fits to the experimental curve. }
\end{figure}

Figures~\ref{Figure 3}(a) and (c) show room-temperature ODMR spectra measured at a distance of $20 \, \mathrm{\mu m}$ from and exactly at the AlN/6H-SiC heterointerface, respectively. One can clearly observe two peaks, labeled as V1/V3 and V2, ascribed to the $\mathrm{V_{Si}}$ centers with different crystallographic environment \cite{PhysRevB.67.125207}. We note that though V1/V3 can be associated with different $\mathrm{V_{Si}}$ centers, they cannot be distinguished in the ODMR spectra. The spectral position of the ODMR peak in zero magnetic field is determined by the zero-filed splitting constant $\nu = 2|D|/h$ of the corresponding $\mathrm{V_{Si}}$ center. To find the ODMR peak positions and determine $2D$ for each $\mathrm{V_{Si}}$, we performed fits with a Gaussian, as shown by the solid lines in Figs.~\ref{Figure 3}(a) and (c). Figure~\ref{Figure 3}(b) shows $| 2D |$ for the V1/V3 and V2 spin centers as a function of the distance from the heterointerface.  Similarly to the Raman peal positions in Fig.~\ref{Figure 2}(b), these dependencies are well fitted with a Gaussian. 
 
At the heterointerface, the V1/V3 ODMR peak is shifted towards lower frequencies, while that corresponding to the V2 center is shifted to higher frequencies in comparison to a stress-free 6H-SiC crystal. The zero-field splitting constants for these spin centers have opposite signs \cite{PhysRevB.98.195204}, it is negative for V1/V3 and positive for V2 (Fig.~\ref{Figure 1}). 

Given that the 6H-SiC crystal is compressively stressed at the heterointerface (i.e., the stress components $\sigma_{\|}$ and $\sigma_{\perp}$ are negative), the observed shifts of the ODMR lines correspond to a positive deformation-induced correction $\delta D$. Using Eq. (\ref{Variation}) and the value of the ODMR shift at the heterointerface, we determine the spin-deformation interaction constants to $\Xi= 0.75 \, \mathrm{GHz}$ for the V1/V3 centers and $\Xi= 0.5 \, \mathrm{GHz}$ for the V2 centers. 

Summarizing, we have measured the effect of static mechanical deformation on the spin properties of the $\mathrm{V_{Si}}$ centers in 6H-SiC and determined the spin-deformation interaction constants. Our results suggest an additional broadening mechanism of the ODMR lines in ensembles of the $\mathrm{V_{Si}}$ centers in SiC and the presence of a set of different spin packets \cite{Soltamov:2019hr}. Though the observed stress-induced shift of the ODMR lines is relatively small, it can be drastically increased using the piezoelectric effect in AlN. Our findings suggest that AlN/SiC heterostructures can allow fine-tuning of the $\mathrm{V_{Si}}$ spin properties and, therefore, are highly promising for quantum applications. 

\begin{acknowledgments}
The authors thank S.A.Tarasenko for fruitful discussions. The authors thank A. Kunz for TEM specimen preparation and Y. Berenc{\'e}n for the Raman test measurements before the preparation.

This study is supported by the Russian Science Foundation (project No.19-72-00154). SIMS measurements were performed using the facility of the Center of Multi-User Equipment  "Material Science and Diagnostics for Advanced Technologies" (Ioffe Institute, Russia), supported by the Russian Ministry of Science (The Agreement ID RFMEFI62119X0021).  The use of the HZDR Ion Beam Center TEM facilities and the funding of TEM Talos by the German Federal Ministry of Education of Research (BMBF), Grant No. 03SF0451, in the framework of HEMCP are acknowledged. 
\end{acknowledgments}
\nocite{*}

%

\end{document}